\documentclass[12pt]{article}
\usepackage[compress]{cite}
\usepackage{hyperref}
\hypersetup{colorlinks,urlcolor=blue,citecolor=blue,linkcolor=blue,filecolor=black}
\usepackage{graphicx}
\usepackage{amsmath}
\usepackage{bm}
\usepackage{amssymb}
\usepackage{footmisc}
\usepackage{slashed}

\newcommand{\z}[1]{\zeta_{#1}}

\newcommand{\M}{n}

\begin{document}

\title{Fermionic contribution to the anomalous dimension\\[2mm] of twist-2 operators in $\mathcal{N}=4$ SYM theory,\\[2mm] critical indices and integrability}

\author{\sc V.N.~Velizhanin\\[5mm]
\it Theoretical Physics Division\\
\it NRC ``Kurchatov Institute''\\
\it Petersburg Nuclear Physics Institute\\
\it Orlova Roscha, Gatchina\\
\it 188300 St.~Petersburg, Russia\\[2mm]
\it velizh@thd.pnpi.spb.ru
}

\date{}

\maketitle

\begin{abstract}
We compute the contribution to the anomalous dimension of the twist-2 operators in $\mathcal{N}=4$ SYM theory, which is proportional to the number of fermion loops inside Feynman diagrams or, formally, to the number of fermions. The result was obtained by the method based on the calculation of critical indices at the critical point by analogy with previous similar computations in scalar theories and in QCD. The obtained result is much simpler with compare to analogous results in QCD and almost satisfies the maximal transcedentality principle. A possible relation between the obtained result and integrability is discussed.
\end{abstract}

\newpage

The anomalous dimension of twist-2 operators is intensively studied in $\mathcal{N}=4$ SYM theory due to the relation with integrability in the framework of AdS/CFT-correspondence\cite{Maldacena:1997re,Gubser:1998bc,Witten:1998qj}. A number of exact results were obtained with the help of integrability, such as the all-loop asymptotic Bethe ansatz~\cite{Minahan:2002ve,Beisert:2003tq,Beisert:2003yb,Bena:2003wd,Kazakov:2004qf,Beisert:2004hm,Arutyunov:2004vx,Staudacher:2004tk,Beisert:2005fw,Janik:2006dc,Eden:2006rx,Beisert:2006ez,Beisert:2010jr}, the TBA-equations and the Y-system~\cite{Arutyunov:2009zu,Gromov:2009tv,Bombardelli:2009ns,Gromov:2009bc,Arutyunov:2009ur}, and finally the quantum spectral curve approach~\cite{Gromov:2013pga,Gromov:2014caa}, which solve the spectral problem for composite operators in $\mathcal{N}=4$ SYM theory.

From the point of view of applicability in realistic models, first of all QCD, the most interesting results obtained in $\mathcal{N}=4$ SYM theory give the most complicated (more precisely, the most transcendental) contribution according to the maximal transcendentality principle~\cite{Kotikov:2002ab}.
In fact, the maximal transcendentality principle was proposed to obtain the results in $\mathcal{N}=4$ SYM theory from the available results in QCD. Initially, an indication of this was obtained during the computations of the eigenvalues of the Balitsky-Fadin-Kuraev-Lipatov (BFKL) equations~\cite{Lipatov:1976zz,Kuraev:1977fs,Balitsky:1978ic} in $\mathcal{N}=4$ SYM theory~\cite{Kotikov:2000pm} based on the original computations in QCD~\cite{Fadin:1998py}. It turned out that if we assign to each term in the general expression for this quantity its transcedentality, which is the sum of absolute values of indices of multiple zeta-values, that are obtained from the function as its argument tends to infinity, then the result in the given order of perturbative theory will contain only functions with maximal transcedentality, equal to $2\ell-1$ for the $\ell$-th order. To prove this assumption, we calculated the two-loop anomalous dimension of twist-2 operators in $\mathcal{N}=4$ SYM theory~\cite{Kotikov:2003fb} and found that it actually holds if we put into the corresponding QCD results~\cite{Gross:1973ju,Georgi:1951sr,Floratos:1977au,GonzalezArroyo:1979df} the following identifications for Casimir operators: $C_F=N_c$ and $C_A=N_c$.

However, in QCD there are also some exact (all-loop) results for twist-2 operators. The most famous result is the eigenvalue of the BFKL equation in the leading and next-to-leading logarithmic approximations~\cite{Lipatov:1976zz,Kuraev:1977fs,Balitsky:1978ic,Fadin:1998py}, from which one can obtain an expression for the anomalous dimension of the gluon operator in any order of perturbative theory near $j=1+\omega$, where it has poles in $\omega$.
Another similar exact result is the double-logarithmic equation for the flavour non-singlet quark twist-2 operator~\cite{Kirschner:1982qf,Kirschner:1983di,Velizhanin:2014dia}, which gives information about the corresponding anomalous dimension in any order of perturbative theory near $j=0+\omega$, where it has poles in $\omega$.

The most interesting all-loop result related to the subject of this paper is the expansion of the non-singlet anomalous dimension in the number of quark flavors or the number of quark loops inside the Feynman diagrams. This result was obtained from the calculation of the critical indices at the critical point within dimensional regularization in Ref.~\cite{Gracey:1994nn}, following the methods proposed in Refs.~\cite{Vasiliev:1981yc,Vasiliev:1981dg}.

In this paper, we compute the contribution proportional to the number of fermionic loops to the anomalous dimension of the special twist-2 operator in $\mathcal{N}=4$ SYM theory, which is similar to the flavor non-singlet operator in QCD. The operator we will consider belong to the representation $\bm{15}$ of the $SU(4)$ group of internal symmetry of $\mathcal{N}=4$ SYM theory and can formally be written in two-componet notation as~\cite{Belitsky:2003sh}
\begin{eqnarray}
\left[{\widetilde{\mathcal{O}}}_{jl}^{qq,15}\right]_A{}^B&=&\sigma_j \mathrm{tr} 
\left[P_{15}\right]^{BC}_{AD}
\bar{\lambda}_{{\dot{\alpha}}C}
\sigma^{{+}\,\dot{\alpha}\beta}
(i\partial^+)^lC^{3/2}_{j}\left(
\overleftrightarrow{\mathcal{D}}^+/\partial^+\right)\lambda^D_\beta\,,\\
\left[P_{15}\right]^{BC}_{AD}&=&\delta^C_A\delta^B_D-\frac{1}{4}\delta^B_A\delta^C_D\,,
\qquad\qquad \sigma_j\equiv 1-(-1)^j,
\end{eqnarray}
where $C^{3/2}_j$ is the Gegenbauer polynomial. In the notation close to usual QCD form it can be written as
\begin{equation}
\widetilde{\mathcal{O}}_{{15}}^{qq}(\M)=
\bar\psi^I\,\gamma_{\mu_1}\!\gamma_5\,\mathcal{D}_{\mu_2}\ldots\mathcal{D}_{\mu_\M}\psi^K
\end{equation}
for even $\M$.

\begin{figure}
\begin{center}
\includegraphics[trim = 0mm 0mm 0mm 0mm, clip,scale=.75,angle=0]{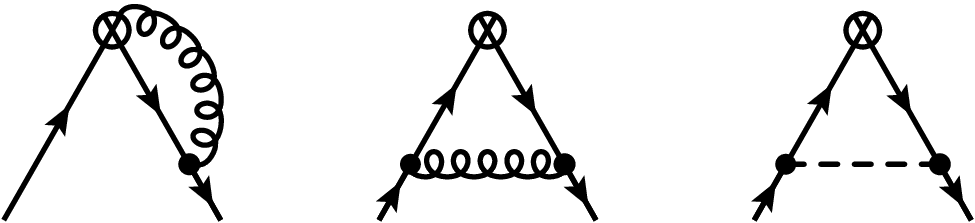}\\
(a)\hspace*{42mm}(b)\hspace*{42mm}(c)\hspace*{0mm}
\caption{One-loop diagrams contributing to anomalous dimension of $\widetilde{\mathcal{O}}_{15}^{qq}$.}
\label{Fig:1L}
\end{center}
\end{figure}

In the leading order, there are three diagrams Fig.\ref{Fig:1L} that contribute to the anomalous dimension of this operator. 
The diagram in Fig.\ref{Fig:1L}.a is the same as in QCD~\cite{Gross:1973ju} and together with its mirror image gives
\begin{equation}
\gamma^{(0)}_{\mathrm{Fig.\ref{Fig:1L}.a+m.d.}}=2\times\big(2\,S_1(\M)-2\big)\,.\label{1La}
\end{equation}
The diagram in Fig.\ref{Fig:1L}.b is also the same as in QCD~\cite{Gross:1973ju} with the following result
\begin{equation}
\gamma^{(0)}_{\mathrm{Fig.\ref{Fig:1L}.b}}=-\frac{2}{\M(\M+1)}\,.\label{1Lb}
\end{equation}
The diagram in Fig.\ref{Fig:1L}.c contains scalar or pseudoscalar fields and the result is the same in both cases, and the overall result is
\begin{equation}
\gamma^{(0)}_{\mathrm{Fig.\ref{Fig:1L}.c}}=2\times\frac{1}{\M(\M+1)}\,.\label{1Lc}
\end{equation}
Putting it all together, we get
\begin{equation}
\gamma^{(0)}_{\widetilde{\mathcal{O}}_{15}^{qq}}(\M)=4\,S_1(\M)\,,\label{1L}
\end{equation}
where the constant in eq.(\ref{1La}) cancels out when the external fermionic field is renormalised.

In what follows, we compute the contribution from insertions of fermion and, in general, scalar loops into gauge and scalar propagators in all order of perturbative theory. Our task is to study the maximal transcedentality principle for the contribution proportional to the number of fermions in order to have additional information that can be used for the reconstruction of the corresponding result in QCD.

For such calculations we use the method proposed in Refs.~\cite{Vasiliev:1981yc,Vasiliev:1981dg} and applied to the computations of the anomalous dimension of non-singlet twist-2 operator in QCD in Ref.~\cite{Gracey:1994nn}. The main idea is that we have to compute the diagrams in Fig.\ref{Fig:1L} with propagators with critical exponents related with critical indices, which can be calculated at the critical point through the expansion in the number of quark (scalar) flavors or the number of quark (scalar) loops.
So, we have to use propagators in the following form~\cite{Vasiliev:1981yc,Vasiliev:1981dg,Gracey:1991kq,Gracey:1992ns,Gracey:1993sn,Gracey:1993ua,Gracey:1994nn}
\begin{eqnarray}
&&
\psi=\frac{A\,\slashed{k}}{\big(k^2)^{\mu-\alpha}}\,,\qquad
\mathcal{A}_{\mu\nu}=\frac{B\, g_{\mu\nu}}{\big(k^2)^{\mu-\beta}}
\,,\qquad
\phi=\frac{C\, }{\big(k^2)^{\mu-\gamma}}\,,\label{PropRul}
\end{eqnarray}
where $A$, $B$ and $C$ are momentum independent amplitudes, $\mu=2-\epsilon$ and $\alpha$, $\beta$ and $\gamma$ are the critical exponents. 
%
%
These amplitudes or their combinations can be fixed by considering the skeleton Dyson equations with dressed propagators~\cite{Vasiliev:1981yc,Vasiliev:1981dg}, which can be represented graphically in the case of gauge and scalar propagators with insertions of fermionic fields~as~\cite{Gracey:1991kq,Gracey:1992ns,Gracey:1993sn,Gracey:1993ua,Gracey:1994nn}
\begin{eqnarray}
0&=&\mathcal{A}_{\mu\nu}^{-1}\ +
\label{Amunu}\\[-12mm]
&&\hspace*{15mm}
\scalebox{-1}[1]{\includegraphics[trim = 0mm 0mm 70mm 80mm, clip,height=1cm,angle=180]{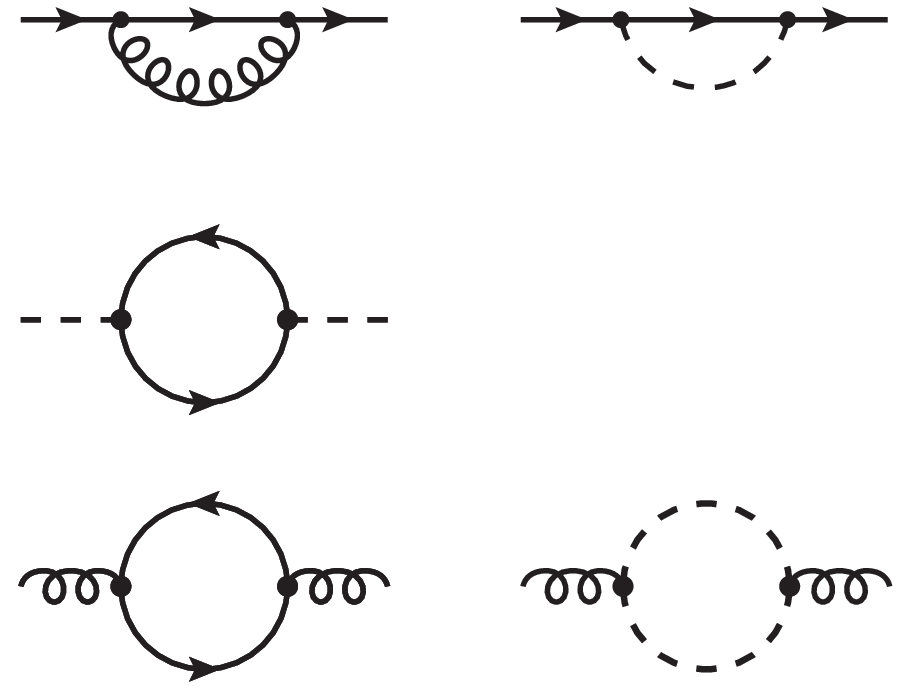}}\nonumber\\[1mm]
0&=&\phi^{-1}\ +
\label{phi}\\[-12mm]
&&\hspace*{15mm}
\scalebox{-1}[1]{\includegraphics[trim = 0mm 44mm 80mm 35mm, clip,height=1cm,angle=180]{QSelf2.eps}}\nonumber
\end{eqnarray}
Strictly speaking, eq.(\ref{Amunu}) is correct only for the gauge field propagator in the Landau gauge $\mathcal{A}_{\mu\nu}=(g_{\mu\nu}-k_\mu k_\nu/k^2)/(k^2)^{\mu-\beta}$, since only the transverse part of the gauge field propagator ensures the correct behaviour of the computed quantities~\cite{Ciuchini:1999cv,Ciuchini:1999wy}, but for the computations of the gauge-invariant quantities, such as anomalous dimension, for the leading $N_f$ contribution one can choose for simplicity the Feynman gauge, as was done in~\cite{Gracey:1994nn}.
After substituting the propagators from eq.(\ref{PropRul}) we obtain the sum of general one-loop integrals, which can be computed using the well-known Tkachov formula~\cite{Tkachov:1981wb}
\begin{eqnarray}
\int\frac{d^D P}{(2\pi)^D}\frac{\mathcal{P}(P)}{P^{2\alpha}(P-Q)^{2\beta}}&=&\nonumber\\
&&\hspace*{-30mm}
\frac{1}{(4\pi)^2}\,(Q^2)^{D/2-\alpha-\beta}\sum_{\sigma\geq 0}^{[n/2]}
G(\alpha,\beta,n,\sigma)\,Q^{2\sigma}
\left\{\frac{1}{\sigma !}\left(\frac{\square}{4}\right)^\sigma\!\mathcal{P}_n(P) \right\}_{P=Q}\label{Tkachov1L}
\end{eqnarray}
where
\begin{equation}
\mathcal{P}_n(P)=P_{\mu_1}P_{\mu_2}\cdots P_{\mu_n}\,,\qquad \qquad\qquad\qquad
\square= \frac{\partial^2}{\partial P_\mu \partial P^\mu}\,,
\end{equation}
$D$ is the dimension of space-time $D=4-2\epsilon$ and $G$ can be expressed in terms of $\Gamma$-functions:
\begin{equation}
G(\alpha,\beta,n,\sigma)=(4\pi)^\epsilon\,
\frac{\Gamma(\alpha+\beta-\sigma-D/2)\,\Gamma(D/2-\alpha+n-\sigma)\,\Gamma(D/2-\beta+\sigma)}{\Gamma(\alpha)\,\Gamma(\beta)\,\Gamma(D-\alpha-\beta+n)}\,.
\end{equation}
We use the Feynman rules from Refs.~\cite{Brink:1976bc,Gliozzi:1976qd}, which we used for all our previous computations. The calculations of eqs.(\ref{Amunu}) and~(\ref{phi}) after substitution of expressions for critical indices in the leading approximation $\alpha=\mu-1$ and $\beta=1$ fixes the combinations of amplitudes $A^2 B$ and $A^2 C$ as~\cite{Gracey:1991kq,Gracey:1992ns,Gracey:1993sn,Gracey:1993ua,Gracey:1994nn}
\begin{eqnarray}
A^2B&=&\frac{\Gamma[2\mu]}{2\,N_f\,\Gamma[2-\mu]\,\Gamma[\mu]^2}\,,\label{A2B}\\
A^2C&=&\frac{\Gamma[2\mu-1]}{N_f\,\Gamma[1-\mu]\,\Gamma[\mu]^2}\,.\label{A2C}
\end{eqnarray}
These results will be used to calculate the diagrams in Fig.\ref{Fig:1L} using the expressions for the propagators from eq.(\ref{PropRul}).

We now turn to the calculation of the diagrams in Fig.\ref{Fig:1L} using the expressions for critical propagators from eq.(\ref{PropRul}). However, if we compute the obtained integrals and expand them after the substitution of the critical indices $\alpha=\mu-1+\eta/2$ and $\beta=1-\eta-\chi_{\mathcal{A}}$ and  $\gamma=1-\eta-\chi_{\phi}$, where $\eta$ is the anomalous dimension of the fermions and $\chi_{\mathcal{A}}$ and $\chi_{\phi}$ are the anomalous dimensions of the respective vertices\footnote{In the leading $N_f$ order they do not give contributions into result, see Refs.~\cite{Gracey:1993sn,Gracey:1993ua,Gracey:1994nn}}, we will find that the integral diverges and therefore requires regularisation. This can be done by shifting the exponents of the gauge and scalar fields by an infinitesimal quantity $\delta$, $\beta\to\beta-\delta$~\cite{Vasiliev:1975mq,Gracey:1991kq,Gracey:1993ua,Gracey:1994nn}. With such regularisation, all diagrams in Fig.\ref{Fig:1L} will have the following formal structure after expanding near $\delta\to0$~\cite{Vasiliev:1975mq,Vasiliev:1983jlo,Gracey:1993sn,Gracey:1994nn}
\begin{equation}
\frac{P}{\delta}+Q+R\,\ln p^2+\mathcal{O}(\delta),\label{deltaExp}
\end{equation}
where $P$, $Q$ and $R$ depend on $\mu$ and $p$ is the external momentum following through the fermions. The simple pole is absorbed by conventional (critical point) renormalisation~\cite{Vasiliev:1975mq,Vasiliev:1983jlo}. The $\delta$-finite Green's function must resumm to the structure $(p^2)^{\gamma^{(n)}_{\mathcal{O}}/2}$~\cite{Vasiliev:1975mq,Vasiliev:1983jlo}, where ${\gamma^{(n)}_{\mathcal{O}}}$ is related to the exponent we are interested in, $\eta_1^{(n)}$, through~\cite{Vasiliev:1975mq,Vasiliev:1983jlo,Gracey:1993sn,Gracey:1994nn}:
\begin{equation}
\eta^{(n)}=\eta+{\gamma^{(n)}_{\mathcal{O}}}
\end{equation}
and is equal to $R$ from eq.(\ref{deltaExp}) or $(-P)$ in the leading $N_f$ order.

The diagrams Fig.\ref{Fig:1L}.a and Fig.\ref{Fig:1L}.b are the same as in QCD with the following already computed results~\cite{Gracey:1994nn}:
\begin{eqnarray}
\eta^{(n)}_{\mathrm{Fig.1.a}+\mathrm{m.d.}}&=&\frac{\mu(\mu-1)\Gamma[2\mu]}{\Gamma[\mu]^2\Gamma[\mu+1]\Gamma[2-\mu]N_f}\sum_{j=2}^n\frac{1}{\mu+j-2}\ ,\label{ResFig1aG}
\\
\eta^{(n)}_{\mathrm{Fig.1.b}}&=&
-\frac{\mu(\mu-1)^3\Gamma[2\mu]}{2\,(\mu+n-1)(\mu+n-2)\Gamma[\mu]^2\Gamma[\mu+1]\Gamma[2-\mu]N_f}\,.
\label{ResFig1bG}
\end{eqnarray}
Expanding the sum of these results in $\epsilon$ in the critical point $\hat{a}_s$ where the one-loop QCD $\beta$-function is equal to zero in the leading $N_f$ order ($a_s=\alpha_s/(4\pi)$)
\begin{equation}
\beta(a_s)=-\epsilon a_s+\bigg(\frac{2}{3}N_f-\frac{11}{3}C_A\bigg)a_s^2+\mathcal{O}(a_s^3)\,,
\quad \hat{a}_s=\frac{3\,\epsilon}{2\,N_f}\,,\quad \epsilon=\frac{2\,\hat{a}_s\,N_f}{3}\,,
\end{equation}
we obtain the results for the leading $N_f$ contribution~\cite{Gracey:1994nn}, which coincide in the first four orders of perturbative theory with the corresponding results from Refs.~\cite{Moch:2002sn,Moch:2004pa,Davies:2016jie}.
Note that in the Landau gauge, only eq.(\ref{ResFig1bG}) gets the constant $(-1/2)$ multiplied by $n$-independent factor from eq.(\ref{ResFig1bG}).

The diagram Fig.\ref{Fig:1L}.c contains the scalar field and together with the result for $A^2C$ from eq.(\ref{A2C}) gives
\begin{equation}
\eta^{(n)}_{\mathrm{Fig.1.c}}=\frac{\mu(\mu-1)^3\Gamma[2\mu-1]}{(\mu+n-1)(\mu+n-2)\Gamma[\mu]^2\Gamma[\mu+1]\Gamma[2-\mu]N_f}\,.\label{ResFig1cG}
\end{equation}
Moreover, we must take into account the contribution from the so-called $\epsilon$-scalars, that appear in the framework of the dimensional reduction scheme~\cite{Siegel:1979wq,Townsend:1979ha,Sezgin:1979bm,Capper:1979ns}. Their contribution is the same as contribution of the scalars~(\ref{ResFig1cG}), but multiplied by $\epsilon=2-\mu$, which together with eq.(\ref{ResFig1cG}) gives
\begin{equation}
\eta^{(n)}_{\mathrm{Fig.1.c}+\epsilon\text{-}\mathrm{sc}}=\frac{\mu(\mu-1)^4\Gamma[2\mu-1]}{(\mu+n-1)(\mu+n-2)\Gamma[\mu]^2\Gamma[\mu+1]\Gamma[2-\mu]N_f}\,.\label{ResFig1cGepS}
\end{equation}
One can see, that the results (\ref{ResFig1bG}) and (\ref{ResFig1cGepS}) for the diagrams Fig.\ref{Fig:1L}.b and Fig.\ref{Fig:1L}.c have different structures and their sum will produce terms proportional to $(1/n/(n+1))^i$ after expansion in $\epsilon$, that is, in higher orders of perturbative theory, while we expect such terms to be absent, as in the expression for the universal anomalous dimension of twist-2 operators in $\mathcal{N}=4$ SYM theory~\cite{Lipatov:2001fs,Kotikov:2003fb,Kotikov:2004er} (see eq.(\ref{1L})).

\begin{figure}
\begin{center}
\includegraphics[trim = 0mm 0mm 0mm 0mm, clip,scale=.75,angle=0]{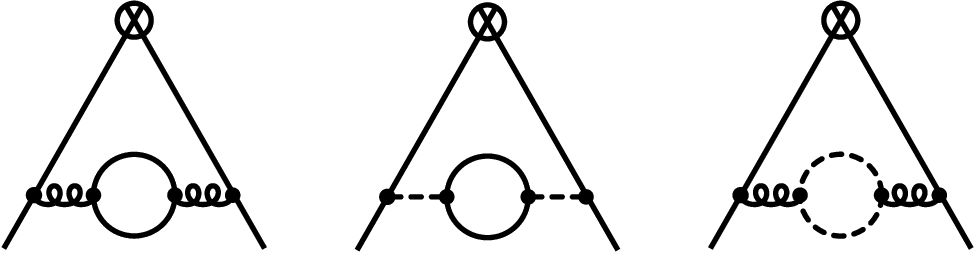}\\
(a)\hspace*{39.5mm}(b)\hspace*{39.5mm}(c)\hspace*{-1mm}
\caption{Two-loop diagrams with fermion and scalar loops insertions.}
\label{Fig:2L}
\end{center}
\end{figure}

To analyse this situation, we used our previous calculations for the two-loop anomalous dimension of twist-2 operators in $\mathcal{N}$=4 SYM theory~\cite{Kotikov:2003fb}. During that calculations we obtained the results for individual diagrams for an arbitrary Lorentz spin of operators. Using these results, we tried to find a combination of diagrams that do not have such terms as $1/n/(n+1)$ in the sum. We found, that it is enough to add to the corresponding two-loops diagrams Fig.\ref{Fig:2L}.a and \ref{Fig:2L}.b, 
whose contributions are respectively
\begin{eqnarray}
\gamma^{(1)}_{\mathrm{Fig.\ref{Fig:2L}.a}}&=&
N_f\left(-\frac{4}{3n^2}
+\frac{4}{3(n+1)^2}
+\frac{44}{9n}
-\frac{44}{9(n+1)}\right)\,,
\label{2La}\\
\gamma^{(1)}_{\mathrm{Fig.\ref{Fig:2L}.b}}&=&
N_f\left(
\frac{2}{n^2}
-\frac{2}{(n+1)^2}
-\frac{6}{n}
+\frac{6}{n+1}\right)
\,,\label{2Lb}
\end{eqnarray}
only one diagram Fig.\ref{Fig:2L}.c with the scalar loop in the gauge field, with the following result
\begin{eqnarray}
\gamma^{(1)}_{\mathrm{Fig.\ref{Fig:2L}.c}}&=&
N_s\left(
-\frac{1}{3n^2}
+\frac{1}{3(n+1)^2}
+\frac{14}{9n}
-\frac{14}{9(n+1)}\right)
\,,\label{2Lc}
\end{eqnarray}
and taking into account the contribution of two $\epsilon$-scalars that must be added within the dimensional reduction scheme~\cite{Siegel:1979wq,Townsend:1979ha,Sezgin:1979bm} and is given by
\begin{eqnarray}
\gamma^{(1)}_{\mathrm{Fig.\ref{Fig:2L}.b}\times2\,\epsilon}&=&
N_f\left(
-\frac{2}{n}
+\frac{2}{n+1}\right)
\,,\label{2Lceps}
\end{eqnarray}
but if we relate the number of scalar fields $N_s$ with the number of fermion fields $N_f$. 
In $\mathcal{N}=1$, $\mathcal{N}=2$ and $\mathcal{N}=4$ SYM theories, the number of scalar fields $N_s$ is related to the number of fermion fields $N_f$ through $N_s=2N_f-2$.
Thus, if we express $N_s$ through $N_f$ in Fig.\ref{Fig:2L}.c, then the sum of all three diagrams in Fig.\ref{Fig:2L} will give zero for the contribution proportional to $N_f$.
This cancellation occurs in higher loops as well, as we checked in three loops by explicit computations with \texttt{MINCER}-package~\cite{Gorishnii:1989gt} for \texttt{FORM}~\cite{Vermaseren:2000nd} the relevant three-loop diagrams, which can be obtained by inserting scalar and fermion loops into the gauge propagators in diagrams Fig.\ref{Fig:2L}.a and~Fig.\ref{Fig:2L}.c and inserting the fermion loops into the scalar propagators in diagram~Fig.\ref{Fig:2L}.b and in three, four and five loops by the direct computations with Tkachov formula~(\ref{Tkachov1L}). We consider only such minimal gauge-invariant subclass of diagrams, since it corresponds to the leading $N_f$ computations in the original paper~\cite{Vasiliev:1981yc} and in QED~\cite{Gracey:1991kq,Gracey:1992ns,Gracey:1993sn} or QCD~\cite{Gracey:1993ua,Gracey:1994nn}, while we want to get results close to the result in QCD. The inclusion of other diagrams will violate the cancellation of $1/n$ terms.

For calculations in the general case with critical exponents, one should additionally consider in eq.(\ref{Amunu}) the diagram with the scalar loop
\begin{eqnarray}
0&=&\mathcal{A}_{\mu\nu}^{-1}\ +
\hspace*{23mm}+
\label{AmunuF}\\[-12mm]
&&\hspace*{15mm}
\scalebox{-1}[1]{\includegraphics[trim = 0mm 0mm 80mm 80mm, clip,height=1cm,angle=180]{QSelf2.eps}}\hspace*{7mm}
\scalebox{-1}[1]{\includegraphics[trim = 80mm 0mm 0mm 80mm, clip,height=1cm,angle=180]{QSelf2.eps}}
\nonumber
\end{eqnarray}
and do not take into account $N_f$ expansion for the scalar propagator here.
The result for eq.(\ref{AmunuF}) has the following form
\begin{eqnarray}
0=1-2\,A^2B\,N_f\,(1-\mu)\,\frac{\Gamma[1-\mu]\,\Gamma[\mu]^2}{\Gamma[2\mu]}
+\,C^2B\,N_s\frac{\Gamma[1-\mu]\,\Gamma[\mu]^2}{2\,\Gamma[2\mu]}
\label{A2BC2B}
\end{eqnarray}
Substituting the expression for $A^2B$ into eq.(\ref{ResFig1bG}) divided by eq.(\ref{A2B})
we fix $C^2B$ and then $A^2B$ demanding the cancellation with eq.(\ref{ResFig1cGepS})
\begin{equation}
{C^2B\,N_s}=
2\,\frac{(\mu-1)\Gamma[2\mu-1]}{\Gamma[2-\mu]\,\Gamma[\mu]^2}\,,\qquad\qquad
{A^2B\,N_f}=
\frac{(\mu-1)\Gamma[2\mu-1]}{\Gamma[2-\mu]\,\Gamma[\mu]^2}\,.
\label{C2BF}
\end{equation}
The results for $A^2B$ and $C^2B$ can be obtained by the direct computations of the critical indices for the fermion and scalar fields and using the Slavnov-Teylor identity between the fermion-fermion-gauge field and scalar-scalar-gauge field vertices, similar to the computations of the critical indices for gluon and ghost fields in QCD in Refs.~\cite{Gracey:1993ua,Ciuchini:1999wy}, but this requires additional consideration of the relevant vertex diagrams.

So, we are left with one diagram Fig.\ref{Fig:1L}.a, which contributes to the critical index for the anomalous dimension. Its contribution is as follows
(with $A^2B$ from eq.(\ref{C2BF}))
\begin{equation}
\eta^{(n)}_{\mathrm{Fig.1.a}+\mathrm{m.d.}}=
2\,A^2B\,\frac{\mu\,(\mu-1)}{\Gamma[\mu+1]}\sum_{j=2}^{n}\frac{1}{\mu+j-2}
=2\,\frac{\mu\,(\mu-1)^2}{N_f\Gamma[\mu+1]}
\frac{\Gamma[2\mu-1]}{\Gamma[2-\mu]\,\Gamma[\mu]^2}
\sum_{j=2}^{n}\frac{1}{\mu+j-2}\,.\label{eta1a}
\end{equation}

Remember now the expression for the one-loop beta function in $\mathcal{N}=1$, $\mathcal{N}=2$ and $\mathcal{N}=4$ SYM theories~\cite{Gross:1973ju,Jones:1974pg}
\begin{equation}
\beta_{\mathrm{SUSY}}(g)=-\epsilon g^2+\bigg(\frac{2}{3}N_f+\frac{1}{6}N_s-\frac{11}{3}\!\bigg)g^4\
\end{equation}
and substituting the above relation between the number of fermionic and scalar fields $N_s=2N_f-2$, we obtain at the critical point $g_c$, where $\beta_{\mathrm{SUSY}}$ equals zero
\begin{equation}
\beta_{\mathrm{SUSY}}(g)=
-\epsilon g^2+
\Big(N_f-4\Big)g^4
,\qquad g_c^2=\frac{\epsilon}{N_f}\,,\qquad \epsilon={N_f} g_c^2\,.
\label{gce}
\end{equation}

Expanding the expression for $\eta^{(n)}$ from eq.(\ref{eta1a}) in $\epsilon$ and using the relation between $\epsilon$ and $g^2$ at the critical point~(\ref{gce}) one can obtain the expression for the anomalous dimension of $\widetilde{\mathcal{O}}^{qq}_{15}$ operator in the number of fermionic loops\footnote{The renormalisation of the external fermions will change the summation over $j$ from $j=2$ to $j=1$.} 
\begin{eqnarray}
\hat{\gamma}^{(n)}_{N_f}&=&
4\,S_1\,g^2
+4(S_2-2\,S_1)\,g^4 N_f
+4(S_3-2\,S_2)\,g^6 N_f^2
+4(S_4-2\,S_3+2\,\z3\,S_1)\,g^8 N_f^3\nonumber\\&&
+4(S_5-2\,S_4+2\,\z3\,S_2-4\,\z3\,S_1+3\,\z4\,S_1)\,g^{10} N_f^4
+\ldots\,\label{gammaNf}
\end{eqnarray}
where the nested harmonic sums $S_{\vec{\bm{a}}}=S_{\vec{\bm{a}}}(n)$ are defined as~\cite{Vermaseren:1998uu}
\begin{equation}
S_i(M)=\sum_{j=1}^M\frac{{\mathrm{sign}}(i)^j}{j^{|i|}}\,,\qquad\quad
S_{i_1,i_2,\ldots,i_k}(M)=\sum_{j=1}^M\frac{{\mathrm{sign}}(i_1)^j}{j^{|i_1|}}S_{i_2,\ldots,i_k}(j)\,.
\end{equation}
The explicit calculations with \texttt{MINCER}~\cite{Gorishnii:1989gt} give the same results for the first two $N_f$ terms in eq.(\ref{gammaNf}).
This expression can be compared with the analogies expression for the leading $N_f$ (the number of fermionic loop) contribution in QCD~\cite{Gracey:1994nn}, which looks like:
\begin{eqnarray}
\hat{\gamma}_{\mathrm{NS}}^{(n)}&=&4\,S_1\,C_F\,\alpha_s
+\left(\frac{8}{3}\,S_2-\frac{40}{9}\,S_1
\right)
C_F\,n_f\,\alpha_s^2
+\left(\frac{16}{9}\,S_3-\frac{80}{27}\,S_2-\frac{16}{27}\,S_1
\right)
C_F\,n_f^2\,\alpha_s^3\nonumber\\
&&+\left(\frac{32}{27}\,S_4-\frac{160}{81}\,S_3-\frac{32}{81}\,S_2-\frac{32}{81}\,S_1
+\frac{64}{27}\,\z3\,S_1
\right)
C_F\,n_f^3\,\alpha_s^4+\ldots\,,\label{nfLeadingQCD}
\end{eqnarray}
where we left only the terms proportional to $S_i$.

Let us now try to relate the obtained result to integrability. Namely, we will try to modify the results of integrability in $\mathcal{N}=4$ SYM theory to include the presented results~(\ref{gammaNf}). According to the relation between the computation of the energy of integrable spin chain and the anomalous dimension of composite operators, we should consider only one-loop diagrams with fermionic and scalar loops insertions into the gauge propagator in Fig.\ref{Fig:1L}.a and Fig.\ref{Fig:1L}.b and fermionin loops insertions into scalar propagator on Fig.\ref{Fig:1L}.c. The most simple way is to consider the solution of the Baxter's equation~\cite{Baxter:1972pc,Baxter:1972wh,Sklyanin:2000sklskl} and modify it in such a way, that the final results match our result~(\ref{gammaNf}).  The Baxter function for twist-2 operators in leading order is expressed through the hypergeometric function
\begin{equation}
Q^{(0)}(u,M)
={}_3 F_2\left(\left. \begin{array}{c}                                                                                                                                  
-M, \ M+1,\ \frac{1}{2}+iu \\                                                                                                                                                 
1,\  1 \end{array}                                                                                                                                                            
\right| 1\right)\!,\ 
i\,\Big(\partial_u Q^{(0)}(u,M)\Big)\Big|^{u=i/2}_{u=\text{-}i/2}=4\,S_1(M)\label{LOBSF}
\end{equation}
and the anomalous dimension in the leading order, equal to $4S_1(M)$, can be found in the standard way by differentiating with respect to $u$ and setting $u$ equals to $i/2$.
Its modifications for higher order were studied in Ref.~\cite{Kotikov:2008pv,Beccaria:2009rw} and we assume that since our two-loop result~(\ref{gammaNf}) gives a part of the full result in the second order, therefore, the next-to-leading order modification of the Baxter function has to contain its in some way. 
We found that the following $\delta$-modification of the Baxter function from Ref.~\cite{Kotikov:2008pv,Beccaria:2009rw}
\begin{equation}
{Q^{(0)}_\delta(u,M)={}_3 F_2\left(\left. \begin{array}{c}
-M, \ M+1+2\delta,\ \frac{1}{2}+iu \\
1+\delta,\  1 \end{array}
\right| 1\right)}
\end{equation}
will give the necessary contributions to the anomalous dimension according to the general formula
\begin{equation}
i\Big(\partial_u
\Big(\partial_\delta^k Q^{(0)}_\delta(u,M)\Big)\Big|_{\delta=0}
\Big)\Big|^{u=i/2}_{u=\text{-}i/2}
= 2\,(-1)^k k!\,S_{k+1}\!\left(\frac{M}{2}\right).
\end{equation}
The modified Baxter function that will give our result~(\ref{gammaNf}) can be written as:
\begin{equation}
Q^{\mathrm{LA}}_{N_f}(u,M) =
\frac{\mu\,(\mu-1)^2}{N_f\Gamma[\mu+1]}
\frac{\Gamma[2\mu-1]}{\Gamma[2-\mu]\,\Gamma[\mu]^2}
\left(
2\,Q^{(0)}_{\,\text{-}\epsilon}(u,2M)
-2\,Q^{(0)}(u,2M)
+\,Q^{(0)}(u,M)
\right)
,
\ \mu=2-\epsilon\,.\label{BFNfLO}
\end{equation}
Expanding the above equation in $\epsilon$~(\ref{gce}) and using eq.(\ref{LOBSF}) we reproduce our result~(\ref{gammaNf}).

However, if we try to go with such minimal modification to the subleading order in $N_f$ expansion, starting with the two-loop Baxter function and its modification similar to eq.(\ref{BFNfLO}) 
\begin{equation}
Q^{(1)}_{N_f}(u,M) =
\frac{\mu\,(\mu-1)^2}{N_f\Gamma[\mu+1]}
\frac{\Gamma[2\mu-1]}{\Gamma[2-\mu]\,\Gamma[\mu]^2}
Q^{(1)}_{\, \epsilon}(u,M)\,,
\label{BFNfNLO}
\end{equation}
with $Q^{(1)}_{\delta}(u,M)$ from Ref.~\cite{Kotikov:2008pv,Beccaria:2009rw} we will not obtain the leading transcedental structures from available results in QCD, which can be extracted from Refs.~\cite{Moch:2002sn,Moch:2004pa,Davies:2016jie} and have the following form:
\begin{eqnarray}
\hat{\gamma}_{\mathrm{NS}}^{(1)}&=&
8C_F(C_A-2 C_F) \Big(
2 S_{1,-2}
-S_{-3}
+ S_3
\Big),\label{ADnf1}\\
\hat{\gamma}_{\mathrm{NS}}^{(2),n_f}&=&
\frac{16}{3}n_f C_F \bigg[(C_A-2 C_F) \Big(
4 S_{1,-2,1}
-2 S_{1,-3}
-2 S_{-3,1}
+ S_{-4}
+3 S_{1,3}
- S_{3,1}
- S_4
\Big) 
\nonumber\\&&
+C_F \Big(
2 S_{1,3}
-2 S_{3,1}
\Big)\bigg],\label{ADnf2}\\
\hat{\gamma}_{\mathrm{NS}}^{(3),n_f^2}&=&\frac{32}{9}n_f^2 C_F\bigg[ (C_A-2 C_F) 
\Big(
8 S_{1,-2,1,1}
-4 S_{1,-2,2}
-4 S_{1,-3,1}
-4 S_{-3,1,1}
+2 S_{1,-4}
\nonumber\\&&
+2 S_{-4,1}
+2 S_{-3,2}
-S_{-5}
+4 S_{1,-2,-2}
-2 S_{-3,-2}
- S_{1,4}
-2 S_{1,3,1}
+ S_{4,1}
+ S_5
\Big)
\nonumber\\&&
-2C_F \Big(
2 S_{1,3,1}
- S_{1,4}
- S_{4,1}
\Big)\bigg].\label{ADnf3}
\end{eqnarray}
Nevertheless, we observed that if we pass from the usual anomalous dimension to the reciprocity-respecting anomalous dimension~\cite{Dokshitzer:2005bf,Dokshitzer:2006nm} the above equations change to
\begin{eqnarray}
\hat{\gamma}_{\mathrm{NS,RR}}^{(1)}&=&C_F(2 C_F-C_A) \Big(
4 \mathbb{B}_{1,2}
-4 \mathbb{B}_{2,1}
\Big),\label{RR1}\\
\hat{\gamma}_{\mathrm{NS,RR}}^{(2),n_f}&=&
n_f C_F\bigg[
(2 C_F-C_A) \Big(
\frac{8}{3} \mathbb{B}_{1,3}
-4 \mathbb{B}_{1,2,1}
+\frac{4}{3} \mathbb{B}_{2,1,1}
\Big) 
+C_F\Big(
\frac{8}{3} \mathbb{B}_{1,2,1}
-\frac{8}{3} \mathbb{B}_{2,1,1}
\Big)\bigg],\quad\ \ \label{RR2}\\
\hat{\gamma}_{\mathrm{NS,RR}}^{(3),n_f^2}&=&
n_f^2C_F \bigg[(2 C_F-C_A) \Big(
\frac{16 }{9}\mathbb{B}_{1,4}
-\frac{16}{9} \mathbb{B}_{1,3,1}
-\frac{8}{9} \mathbb{B}_{2,2,1}
+\frac{8}{9} \mathbb{B}_{1,2,1,1}
\Big)\nonumber\\
&& \qquad\qquad\qquad\qquad+C_F^2 \Big(
\frac{16}{9} \mathbb{B}_{2,2,1}
-\frac{16}{9}\mathbb{B}_{1,2,1,1}
\Big)\bigg],\label{RR3}
\end{eqnarray}
where the binomial harmonic sums $\mathbb{B}_{\vec{\bm{a}}}=\mathbb{B}_{\vec{\bm{a}}}(M)$ are defined as~\cite{Vermaseren:1998uu}
\begin{equation}
\mathbb{B}_{\vec{\bm{a}}}(M)=\sum_{k=1}^M\binom{M}{k}\binom{M+k}{k}S_{\vec{\bm{a}}}(k)\,.
\end{equation}
The first $\mathbb{B}_{\vec{\bm{a}}}$ in eqs.(\ref{RR1})-(\ref{RR3}) that correspond to the terms in eqs.(\ref{ADnf1})-(\ref{ADnf3}) with the maximal number of indices can be generated by the following sequence of the binomial sums:
\begin{equation}
\mathbb{B}_{1,2}\quad\mapsto\quad\mathbb{B}_{1,3}\quad\mapsto\quad\mathbb{B}_{1,4}\,,
\end{equation}
that is, from $\mathbb{B}_{1,2}$ formally differentiated by the last index with exactly the same relative coefficients as for the leading $n_f$ contributions~(\ref{nfLeadingQCD}).
It seems, that the most complicated part for the subleading $N_f$ contribution can be generated using an expression similar to eq.(\ref{ResFig1aG}) but with the usual harmonic sums changed to binomial harmonic sums. Additional information can be obtained from the relevant direct diagrammatic calculations, which are in progress both in QCD and in $\mathcal{N}=4$ SYM theory.

We expect that in higher loops there is some set of diagrams that gives the most transcedental terms to $N_f$ contribution, and this set of diagrams contains QCD diagrams and diagrams in which gauge fields are replaced by scalar fields, canceling the ``simple'' terms, especially with $1/n$, as discussed here. The number of terms in the final result will decrease due to such cancellations. This will be very important for the four-loop computations, because so far the only way to get the full result for the anomalous dimension of twist-2 operators in four loops is to compute the results for the first few fixed moments of operators and reconstruct the general result using number theory, assume that the final results will consist of nested harmonic sums\footnote{Such method, proposed by us in Ref.~\cite{Velizhanin:2010cm}, was successfully applied for the computation at four loops for the non-planar contribution to the universal anomalous dimension in $\mathcal{N}=4$ SYM theory~\cite{Kniehl:2020rip,Kniehl:2021ysp}, $n_f^2$~contribution to the non-singlet anomalous dimension in QCD~\cite{Davies:2016jie} and planar limit of the non-singlet anomalous dimension in QCD~\cite{Moch:2017uml}.}. Any additional information, in particularly, that can be obtained by extending the consideration presented in this paper, will allow us finally to find the general result for $n_f$ contribution to the four-loop anomalous dimension of the non-singlet twist-2 operator in QCD.


 \subsection*{Acknowledgments}

The research was supported by a grant from the Russian Science Foundation No. 22-22-00803, https://rscf.ru/en/project/22-22-00803/.


\begin{thebibliography}{10}

\bibitem{Maldacena:1997re}
J.M.~Maldacena, \emph{{The Large N limit of superconformal field theories and
  supergravity}}, \href{https://doi.org/10.1023/A:1026654312961}{\emph{Adv.
  Theor. Math. Phys.} {\bfseries 2} (1998) 231}
  [\href{https://arxiv.org/abs/hep-th/9711200}{{\ttfamily hep-th/9711200}}].

\bibitem{Gubser:1998bc}
S.S.~Gubser, I.R.~Klebanov and A.M.~Polyakov, \emph{{Gauge theory correlators
  from noncritical string theory}},
  \href{https://doi.org/10.1016/S0370-2693(98)00377-3}{\emph{Phys. Lett. B}
  {\bfseries 428} (1998) 105}
  [\href{https://arxiv.org/abs/hep-th/9802109}{{\ttfamily hep-th/9802109}}].

\bibitem{Witten:1998qj}
E.~Witten, \emph{{Anti-de Sitter space and holography}},
  \href{https://doi.org/10.4310/ATMP.1998.v2.n2.a2}{\emph{Adv. Theor. Math.
  Phys.} {\bfseries 2} (1998) 253}
  [\href{https://arxiv.org/abs/hep-th/9802150}{{\ttfamily hep-th/9802150}}].

\bibitem{Minahan:2002ve}
J.A.~Minahan and K.~Zarembo, \emph{{The Bethe ansatz for N=4 superYang-Mills}},
  \href{https://doi.org/10.1088/1126-6708/2003/03/013}{\emph{JHEP} {\bfseries
  03} (2003) 013} [\href{https://arxiv.org/abs/hep-th/0212208}{{\ttfamily
  hep-th/0212208}}].

\bibitem{Beisert:2003tq}
N.~Beisert, C.~Kristjansen and M.~Staudacher, \emph{{The Dilatation operator of
  conformal N=4 superYang-Mills theory}},
  \href{https://doi.org/10.1016/S0550-3213(03)00406-1}{\emph{Nucl. Phys. B}
  {\bfseries 664} (2003) 131}
  [\href{https://arxiv.org/abs/hep-th/0303060}{{\ttfamily hep-th/0303060}}].

\bibitem{Beisert:2003yb}
N.~Beisert and M.~Staudacher, \emph{{The N=4 SYM integrable super spin chain}},
  \href{https://doi.org/10.1016/j.nuclphysb.2003.08.015}{\emph{Nucl. Phys. B}
  {\bfseries 670} (2003) 439}
  [\href{https://arxiv.org/abs/hep-th/0307042}{{\ttfamily hep-th/0307042}}].

\bibitem{Bena:2003wd}
I.~Bena, J.~Polchinski and R.~Roiban, \emph{{Hidden symmetries of the AdS(5) x
  S**5 superstring}},
  \href{https://doi.org/10.1103/PhysRevD.69.046002}{\emph{Phys. Rev. D}
  {\bfseries 69} (2004) 046002}
  [\href{https://arxiv.org/abs/hep-th/0305116}{{\ttfamily hep-th/0305116}}].

\bibitem{Kazakov:2004qf}
V.A.~Kazakov, A.~Marshakov, J.A.~Minahan and K.~Zarembo,
  \emph{{Classical/quantum integrability in AdS/CFT}},
  \href{https://doi.org/10.1088/1126-6708/2004/05/024}{\emph{JHEP} {\bfseries
  05} (2004) 024} [\href{https://arxiv.org/abs/hep-th/0402207}{{\ttfamily
  hep-th/0402207}}].

\bibitem{Beisert:2004hm}
N.~Beisert, V.~Dippel and M.~Staudacher, \emph{{A Novel long range spin chain
  and planar N=4 super Yang-Mills}},
  \href{https://doi.org/10.1088/1126-6708/2004/07/075}{\emph{JHEP} {\bfseries
  07} (2004) 075} [\href{https://arxiv.org/abs/hep-th/0405001}{{\ttfamily
  hep-th/0405001}}].

\bibitem{Arutyunov:2004vx}
G.~Arutyunov, S.~Frolov and M.~Staudacher, \emph{{Bethe ansatz for quantum
  strings}}, \href{https://doi.org/10.1088/1126-6708/2004/10/016}{\emph{JHEP}
  {\bfseries 10} (2004) 016}
  [\href{https://arxiv.org/abs/hep-th/0406256}{{\ttfamily hep-th/0406256}}].

\bibitem{Staudacher:2004tk}
M.~Staudacher, \emph{{The Factorized S-matrix of CFT/AdS}},
  \href{https://doi.org/10.1088/1126-6708/2005/05/054}{\emph{JHEP} {\bfseries
  05} (2005) 054} [\href{https://arxiv.org/abs/hep-th/0412188}{{\ttfamily
  hep-th/0412188}}].

\bibitem{Beisert:2005fw}
N.~Beisert and M.~Staudacher, \emph{{Long-range psu(2,2|4) Bethe Ansatze for
  gauge theory and strings}},
  \href{https://doi.org/10.1016/j.nuclphysb.2005.06.038}{\emph{Nucl. Phys. B}
  {\bfseries 727} (2005) 1}
  [\href{https://arxiv.org/abs/hep-th/0504190}{{\ttfamily hep-th/0504190}}].

\bibitem{Janik:2006dc}
R.A.~Janik, \emph{{The AdS(5) x S**5 superstring worldsheet S-matrix and
  crossing symmetry}},
  \href{https://doi.org/10.1103/PhysRevD.73.086006}{\emph{Phys. Rev. D}
  {\bfseries 73} (2006) 086006}
  [\href{https://arxiv.org/abs/hep-th/0603038}{{\ttfamily hep-th/0603038}}].

\bibitem{Eden:2006rx}
B.~Eden and M.~Staudacher, \emph{{Integrability and transcendentality}},
  \href{https://doi.org/10.1088/1742-5468/2006/11/P11014}{\emph{J. Stat. Mech.}
  {\bfseries 0611} (2006) P11014}
  [\href{https://arxiv.org/abs/hep-th/0603157}{{\ttfamily hep-th/0603157}}].

\bibitem{Beisert:2006ez}
N.~Beisert, B.~Eden and M.~Staudacher, \emph{{Transcendentality and Crossing}},
  \href{https://doi.org/10.1088/1742-5468/2007/01/P01021}{\emph{J. Stat. Mech.}
  {\bfseries 0701} (2007) P01021}
  [\href{https://arxiv.org/abs/hep-th/0610251}{{\ttfamily hep-th/0610251}}].

\bibitem{Beisert:2010jr}
N.~Beisert et~al., \emph{{Review of AdS/CFT Integrability: An Overview}},
  \href{https://doi.org/10.1007/s11005-011-0529-2}{\emph{Lett. Math. Phys.}
  {\bfseries 99} (2012) 3} [\href{https://arxiv.org/abs/1012.3982}{{\ttfamily
  1012.3982}}].

\bibitem{Arutyunov:2009zu}
G.~Arutyunov and S.~Frolov, \emph{{String hypothesis for the AdS(5) x S**5
  mirror}}, \href{https://doi.org/10.1088/1126-6708/2009/03/152}{\emph{JHEP}
  {\bfseries 03} (2009) 152} [\href{https://arxiv.org/abs/0901.1417}{{\ttfamily
  0901.1417}}].

\bibitem{Gromov:2009tv}
N.~Gromov, V.~Kazakov and P.~Vieira, \emph{{Exact spectrum of anomalous
  dimensions of planar N=4 supersymmetric Yang-Mills theory}},
  \href{https://doi.org/10.1103/PhysRevLett.103.131601}{\emph{Phys. Rev. Lett.}
  {\bfseries 103} (2009) 131601}
  [\href{https://arxiv.org/abs/0901.3753}{{\ttfamily 0901.3753}}].

\bibitem{Bombardelli:2009ns}
D.~Bombardelli, D.~Fioravanti and R.~Tateo, \emph{{Thermodynamic Bethe Ansatz
  for planar AdS/CFT: A Proposal}},
  \href{https://doi.org/10.1088/1751-8113/42/37/375401}{\emph{J. Phys. A}
  {\bfseries 42} (2009) 375401}
  [\href{https://arxiv.org/abs/0902.3930}{{\ttfamily 0902.3930}}].

\bibitem{Gromov:2009bc}
N.~Gromov, V.~Kazakov, A.~Kozak and P.~Vieira, \emph{{Exact spectrum of
  anomalous dimensions of planar N = 4 supersymmetric Yang-Mills theory: TBA
  and excited states}},
  \href{https://doi.org/10.1007/s11005-010-0374-8}{\emph{Lett. Math. Phys.}
  {\bfseries 91} (2010) 265} [\href{https://arxiv.org/abs/0902.4458}{{\ttfamily
  0902.4458}}].

\bibitem{Arutyunov:2009ur}
G.~Arutyunov and S.~Frolov, \emph{{Thermodynamic Bethe Ansatz for the AdS(5) x
  S(5) Mirror Model}},
  \href{https://doi.org/10.1088/1126-6708/2009/05/068}{\emph{JHEP} {\bfseries
  05} (2009) 068} [\href{https://arxiv.org/abs/0903.0141}{{\ttfamily
  0903.0141}}].

\bibitem{Gromov:2013pga}
N.~Gromov, V.~Kazakov, S.~Leurent and D.~Volin, \emph{{Quantum spectral curve
  for planar $\mathcal{N} = 4$ Super-Yang-Mills theory}},
  \href{https://doi.org/10.1103/PhysRevLett.112.011602}{\emph{Phys. Rev. Lett.}
  {\bfseries 112} (2014) 011602}
  [\href{https://arxiv.org/abs/1305.1939}{{\ttfamily 1305.1939}}].

\bibitem{Gromov:2014caa}
N.~Gromov, V.~Kazakov, S.~Leurent and D.~Volin, \emph{{Quantum spectral curve
  for arbitrary state/operator in AdS$_{5}$/CFT$_{4}$}},
  \href{https://doi.org/10.1007/JHEP09(2015)187}{\emph{JHEP} {\bfseries 09}
  (2015) 187} [\href{https://arxiv.org/abs/1405.4857}{{\ttfamily 1405.4857}}].

\bibitem{Kotikov:2002ab}
A.V.~Kotikov and L.N.~Lipatov, \emph{{DGLAP and BFKL equations in the $N=4$
  supersymmetric gauge theory}},
  \href{https://doi.org/10.1016/S0550-3213(03)00264-5}{\emph{Nucl. Phys. B}
  {\bfseries 661} (2003) 19}
  [\href{https://arxiv.org/abs/hep-ph/0208220}{{\ttfamily hep-ph/0208220}}].

\bibitem{Lipatov:1976zz}
L.N.~Lipatov, \emph{{Reggeization of the vector meson and the vacuum
  singularity in nonabelian gauge theories}}, {\emph{Sov. J. Nucl. Phys.}
  {\bfseries 23} (1976) 338}.

\bibitem{Kuraev:1977fs}
E.A.~Kuraev, L.N.~Lipatov and V.S.~Fadin, \emph{{The Pomeranchuk singularity in
  nonabelian gauge theories}}, {\emph{Sov. Phys. JETP} {\bfseries 45} (1977)
  199}.

\bibitem{Balitsky:1978ic}
I.I.~Balitsky and L.N.~Lipatov, \emph{{The Pomeranchuk singularity in quantum
  chromodynamics}}, {\emph{Sov. J. Nucl. Phys.} {\bfseries 28} (1978) 822}.

\bibitem{Kotikov:2000pm}
A.V.~Kotikov and L.N.~Lipatov, \emph{{NLO corrections to the BFKL equation in
  QCD and in supersymmetric gauge theories}},
  \href{https://doi.org/10.1016/S0550-3213(00)00329-1}{\emph{Nucl. Phys. B}
  {\bfseries 582} (2000) 19}
  [\href{https://arxiv.org/abs/hep-ph/0004008}{{\ttfamily hep-ph/0004008}}].

\bibitem{Fadin:1998py}
V.S.~Fadin and L.N.~Lipatov, \emph{{BFKL pomeron in the next-to-leading
  approximation}},
  \href{https://doi.org/10.1016/S0370-2693(98)00473-0}{\emph{Phys. Lett. B}
  {\bfseries 429} (1998) 127}
  [\href{https://arxiv.org/abs/hep-ph/9802290}{{\ttfamily hep-ph/9802290}}].

\bibitem{Kotikov:2003fb}
A.V.~Kotikov, L.N.~Lipatov and V.N.~Velizhanin, \emph{{Anomalous dimensions of
  Wilson operators in N=4 SYM theory}},
  \href{https://doi.org/10.1016/S0370-2693(03)00184-9}{\emph{Phys. Lett. B}
  {\bfseries 557} (2003) 114}
  [\href{https://arxiv.org/abs/hep-ph/0301021}{{\ttfamily hep-ph/0301021}}].

\bibitem{Gross:1973ju}
D.J.~Gross and F.~Wilczek, \emph{{Asymptotically free gauge theories - I}},
  \href{https://doi.org/10.1103/PhysRevD.8.3633}{\emph{Phys. Rev. D} {\bfseries
  8} (1973) 3633}.

\bibitem{Georgi:1951sr}
H.~Georgi and H.D.~Politzer, \emph{{Electroproduction scaling in an
  asymptotically free theory of strong interactions}},
  \href{https://doi.org/10.1103/PhysRevD.9.416}{\emph{Phys. Rev. D} {\bfseries
  9} (1974) 416}.

\bibitem{Floratos:1977au}
E.G.~Floratos, D.A.~Ross and C.T.~Sachrajda, \emph{{Higher order effects in
  asymptotically free gauge theories: the anomalous dimensions of Wilson
  operators}}, \href{https://doi.org/10.1016/0550-3213(77)90020-7}{\emph{Nucl.
  Phys. B} {\bfseries 129} (1977) 66}.

\bibitem{GonzalezArroyo:1979df}
A.~Gonzalez-Arroyo, C.~Lopez and F.J.~Yndurain, \emph{{Second order
  contributions to the structure functions in deep inelastic scattering. 1.
  Theoretical calculations}},
  \href{https://doi.org/10.1016/0550-3213(79)90596-0}{\emph{Nucl. Phys. B}
  {\bfseries 153} (1979) 161}.

\bibitem{Kirschner:1982qf}
R.~Kirschner and L.N.~Lipatov, \emph{{Double logarithmic asymptotics of quark
  scattering amplitudes with flavor exchange}},
  \href{https://doi.org/10.1103/PhysRevD.26.1202}{\emph{Phys. Rev. D}
  {\bfseries 26} (1982) 1202}.

\bibitem{Kirschner:1983di}
R.~Kirschner and L.n.~Lipatov, \emph{{Double logarithmic asymptotics and Regge
  singularities of quark amplitudes with flavor exchange}},
  \href{https://doi.org/10.1016/0550-3213(83)90178-5}{\emph{Nucl. Phys. B}
  {\bfseries 213} (1983) 122}.

\bibitem{Velizhanin:2014dia}
V.N.~Velizhanin, \emph{{Generalised double-logarithmic equation in QCD}},
  \href{https://doi.org/10.1142/S0217732317502133}{\emph{Mod. Phys. Lett. A}
  {\bfseries 32} (2017) 1750213}
  [\href{https://arxiv.org/abs/1412.7143}{{\ttfamily 1412.7143}}].

\bibitem{Gracey:1994nn}
J.A.~Gracey, \emph{{Anomalous dimension of nonsinglet Wilson operators at O (1
  / N(f)) in deep inelastic scattering}},
  \href{https://doi.org/10.1016/0370-2693(94)90502-9}{\emph{Phys. Lett. B}
  {\bfseries 322} (1994) 141}
  [\href{https://arxiv.org/abs/hep-ph/9401214}{{\ttfamily hep-ph/9401214}}].

\bibitem{Vasiliev:1981yc}
A.N.~Vasiliev, Y.M.~Pismak and Y.R.~Khonkonen, \emph{{Simple method of
  calculating the critical indices in the 1/$N$ expansion}},
  \href{https://doi.org/10.1007/BF01030844}{\emph{Theor. Math. Phys.}
  {\bfseries 46} (1981) 104}.

\bibitem{Vasiliev:1981dg}
A.N.~Vasiliev, Y.M.~Pismak and Y.R.~Khonkonen, \emph{{1/$N$ expansion:
  calculation of the exponents $\eta$ and Nu in the order 1/$N^2$ for arbitrary
  number of dimensions}},
  \href{https://doi.org/10.1007/BF01019296}{\emph{Theor. Math. Phys.}
  {\bfseries 47} (1981) 465}.

\bibitem{Belitsky:2003sh}
A.V.~Belitsky, S.E.~Derkachov, G.P.~Korchemsky and A.N.~Manashov,
  \emph{{Superconformal operators in N=4 superYang-Mills theory}},
  \href{https://doi.org/10.1103/PhysRevD.70.045021}{\emph{Phys. Rev. D}
  {\bfseries 70} (2004) 045021}
  [\href{https://arxiv.org/abs/hep-th/0311104}{{\ttfamily hep-th/0311104}}].

\bibitem{Gracey:1991kq}
J.A.~Gracey, \emph{{Quantum electrodynamics at O(1/N-f(2)) in arbitrary
  dimensions}}, \href{https://doi.org/10.1142/S0217732392001658}{\emph{Mod.
  Phys. Lett. A} {\bfseries 7} (1992) 1945}.

\bibitem{Gracey:1992ns}
J.A.~Gracey, \emph{{Algorithm for computing the beta function of quantum
  electrodynamics in the large N(f) expansion}},
  \href{https://doi.org/10.1142/S0217751X93000977}{\emph{Int. J. Mod. Phys. A}
  {\bfseries 8} (1993) 2465}
  [\href{https://arxiv.org/abs/hep-th/9301123}{{\ttfamily hep-th/9301123}}].

\bibitem{Gracey:1993sn}
J.A.~Gracey, \emph{{Electron mass anomalous dimension at O(1/(Nf(2)) in quantum
  electrodynamics}},
  \href{https://doi.org/10.1016/0370-2693(93)91017-H}{\emph{Phys. Lett. B}
  {\bfseries 317} (1993) 415}
  [\href{https://arxiv.org/abs/hep-th/9309092}{{\ttfamily hep-th/9309092}}].

\bibitem{Gracey:1993ua}
J.A.~Gracey, \emph{{Quark, gluon and ghost anomalous dimensions at O(1/N(f)) in
  quantum chromodynamics}},
  \href{https://doi.org/10.1016/0370-2693(93)91803-U}{\emph{Phys. Lett. B}
  {\bfseries 318} (1993) 177}
  [\href{https://arxiv.org/abs/hep-th/9310063}{{\ttfamily hep-th/9310063}}].

\bibitem{Ciuchini:1999cv}
M.~Ciuchini, S.E.~Derkachov, J.A.~Gracey and A.N.~Manashov, \emph{{Quark mass
  anomalous dimension at O(1/N(f)**2) in QCD}},
  \href{https://doi.org/10.1016/S0370-2693(99)00573-0}{\emph{Phys. Lett. B}
  {\bfseries 458} (1999) 117}
  [\href{https://arxiv.org/abs/hep-ph/9903410}{{\ttfamily hep-ph/9903410}}].

\bibitem{Ciuchini:1999wy}
M.~Ciuchini, S.E.~Derkachov, J.A.~Gracey and A.N.~Manashov, \emph{{Computation
  of quark mass anomalous dimension at O(1 / N**2(f)) in quantum
  chromodynamics}},
  \href{https://doi.org/10.1016/S0550-3213(00)00209-1}{\emph{Nucl. Phys. B}
  {\bfseries 579} (2000) 56}
  [\href{https://arxiv.org/abs/hep-ph/9912221}{{\ttfamily hep-ph/9912221}}].

\bibitem{Tkachov:1981wb}
F.V.~Tkachov, \emph{{A theorem on analytical calculability of four loop
  renormalization group functions}},
  \href{https://doi.org/10.1016/0370-2693(81)90288-4}{\emph{Phys. Lett. B}
  {\bfseries 100} (1981) 65}.

\bibitem{Brink:1976bc}
L.~Brink, J.H.~Schwarz and J.~Scherk, \emph{{Supersymmetric Yang-Mills
  theories}}, \href{https://doi.org/10.1016/0550-3213(77)90328-5}{\emph{Nucl.
  Phys. B} {\bfseries 121} (1977) 77}.

\bibitem{Gliozzi:1976qd}
F.~Gliozzi, J.~Scherk and D.I.~Olive, \emph{{Supersymmetry, supergravity
  theories and the dual spinor model}},
  \href{https://doi.org/10.1016/0550-3213(77)90206-1}{\emph{Nucl. Phys. B}
  {\bfseries 122} (1977) 253}.

\bibitem{Vasiliev:1975mq}
A.N.~Vasiliev and M.Y.~Nalimov, \emph{{Analog of dimensional regularization for
  calculation of the renormalization group functions in the 1/n expansion for
  arbitrary dimension of space}},
  \href{https://doi.org/10.1007/BF01015800}{\emph{Theor. Math. Phys.}
  {\bfseries 55} (1983) 423}.

\bibitem{Vasiliev:1983jlo}
A.N.~Vasiliev and M.Y.~Nalimov, \emph{{The CP**(n-1) model: calculation of
  anomalous dimensions and the mixing matrices in the order 1/N}},
  \href{https://doi.org/10.1007/BF01027537}{\emph{Theor. Math. Phys.}
  {\bfseries 56} (1983) 643}.

\bibitem{Moch:2002sn}
S.~Moch, J.A.M.~Vermaseren and A.~Vogt, \emph{{Nonsinglet structure functions
  at three loops: Fermionic contributions}},
  \href{https://doi.org/10.1016/S0550-3213(02)00870-2}{\emph{Nucl. Phys. B}
  {\bfseries 646} (2002) 181}
  [\href{https://arxiv.org/abs/hep-ph/0209100}{{\ttfamily hep-ph/0209100}}].

\bibitem{Moch:2004pa}
S.~Moch, J.A.M.~Vermaseren and A.~Vogt, \emph{{The three loop splitting
  functions in QCD: The nonsinglet case}},
  \href{https://doi.org/10.1016/j.nuclphysb.2004.03.030}{\emph{Nucl. Phys. B}
  {\bfseries 688} (2004) 101}
  [\href{https://arxiv.org/abs/hep-ph/0403192}{{\ttfamily hep-ph/0403192}}].

\bibitem{Davies:2016jie}
J.~Davies, A.~Vogt, B.~Ruijl, T.~Ueda and J.A.M.~Vermaseren, \emph{{Large-$n_f$
  contributions to the four-loop splitting functions in QCD}},
  \href{https://doi.org/10.1016/j.nuclphysb.2016.12.012}{\emph{Nucl. Phys. B}
  {\bfseries 915} (2017) 335}
  [\href{https://arxiv.org/abs/1610.07477}{{\ttfamily 1610.07477}}].

\bibitem{Siegel:1979wq}
W.~Siegel, \emph{{Supersymmetric dimensional regularization via dimensional
  reduction}}, \href{https://doi.org/10.1016/0370-2693(79)90282-X}{\emph{Phys.
  Lett. B} {\bfseries 84} (1979) 193}.

\bibitem{Townsend:1979ha}
P.K.~Townsend and P.~van Nieuwenhuizen, \emph{{Dimensional regularization and
  supersymmetry at the two loop level}},
  \href{https://doi.org/10.1103/PhysRevD.20.1832}{\emph{Phys. Rev. D}
  {\bfseries 20} (1979) 1832}.

\bibitem{Sezgin:1979bm}
E.~Sezgin, \emph{{Dimensional regularization and the massive {Wess-Zumino}
  model}}, \href{https://doi.org/10.1016/0550-3213(80)90428-9}{\emph{Nucl.
  Phys. B} {\bfseries 162} (1980) 1}.

\bibitem{Capper:1979ns}
D.M.~Capper, D.R.T.~Jones and P.~van Nieuwenhuizen, \emph{{Regularization by
  Dimensional Reduction of Supersymmetric and Nonsupersymmetric Gauge
  Theories}}, \href{https://doi.org/10.1016/0550-3213(80)90244-8}{\emph{Nucl.
  Phys. B} {\bfseries 167} (1980) 479}.

\bibitem{Lipatov:2001fs}
L.N.~Lipatov, \emph{{Next-to-leading corrections to the BFKL equation and the
  effective action for high energy processes in QCD}},
  \href{https://doi.org/10.1016/S0920-5632(01)01329-9}{\emph{Nucl. Phys. B
  Proc. Suppl.} {\bfseries 99} (2001) 175}.

\bibitem{Kotikov:2004er}
A.V.~Kotikov, L.N.~Lipatov, A.I.~Onishchenko and V.N.~Velizhanin, \emph{{Three
  loop universal anomalous dimension of the Wilson operators in $N=4$ SUSY
  Yang-Mills model}},
  \href{https://doi.org/10.1016/j.physletb.2004.05.078}{\emph{Phys. Lett. B}
  {\bfseries 595} (2004) 521}
  [\href{https://arxiv.org/abs/hep-th/0404092}{{\ttfamily hep-th/0404092}}].

\bibitem{Gorishnii:1989gt}
S.G.~Gorishnii, S.A.~Larin, L.R.~Surguladze and F.V.~Tkachov, \emph{{Mincer:
  program for multiloop calculations in quantum field theory for the
  Schoonschip system}},
  \href{https://doi.org/10.1016/0010-4655(89)90134-3}{\emph{Comput. Phys.
  Commun.} {\bfseries 55} (1989) 381}.

\bibitem{Vermaseren:2000nd}
J.A.M.~Vermaseren, \emph{{New features of FORM}},
  \href{https://arxiv.org/abs/math-ph/0010025}{{\ttfamily math-ph/0010025}}.

\bibitem{Jones:1974pg}
D.R.T.~Jones, \emph{{Asymptotic behavior of supersymmetric Yang-Mills theories
  in the two loop approximation}},
  \href{https://doi.org/10.1016/0550-3213(75)90256-4}{\emph{Nucl. Phys. B}
  {\bfseries 87} (1975) 127}.

\bibitem{Vermaseren:1998uu}
J.A.M.~Vermaseren, \emph{{Harmonic sums, Mellin transforms and integrals}},
  \href{https://doi.org/10.1142/S0217751X99001032}{\emph{Int. J. Mod. Phys. A}
  {\bfseries 14} (1999) 2037}
  [\href{https://arxiv.org/abs/hep-ph/9806280}{{\ttfamily hep-ph/9806280}}].

\bibitem{Baxter:1972pc}
R.J.~Baxter, \emph{{One-dimensional anisotropic Heisenberg chain}},
  \href{https://doi.org/10.1016/0003-4916(72)90270-9}{\emph{Annals Phys.}
  {\bfseries 70} (1972) 323}.

\bibitem{Baxter:1972wh}
R.J.~Baxter, \emph{{Eight vertex model in lattice statistics and
  one-dimensional anisotropic Heisenberg chain. 1. Eigenvectors of the transfer
  matrix and Hamiltonian}},
  \href{https://doi.org/10.1016/0003-4916(73)90441-7}{\emph{Annals Phys.}
  {\bfseries 76} (1973) 48}.

\bibitem{Sklyanin:2000sklskl}
E.K.~Sklyanin, \emph{{B\"acklund transformations and Baxter's Q-operator}},
  \href{https://doi.org/10.1090/crmp/026}{\emph{CRM Proceedings and Lecture
  Notes, Amer. Math. Soc., Providence, RI} {\bfseries 26} (2000) 227}
  [\href{https://arxiv.org/abs/nlin/0009009}{{\ttfamily nlin/0009009}}].

\bibitem{Kotikov:2008pv}
A.V.~Kotikov, A.~Rej and S.~Zieme, \emph{{Analytic three-loop solutions for N=4
  SYM twist operators}},
  \href{https://doi.org/10.1016/j.nuclphysb.2008.12.022}{\emph{Nucl. Phys. B}
  {\bfseries 813} (2009) 460}
  [\href{https://arxiv.org/abs/0810.0691}{{\ttfamily 0810.0691}}].

\bibitem{Beccaria:2009rw}
M.~Beccaria, A.V.~Belitsky, A.V.~Kotikov and S.~Zieme, \emph{{Analytic solution
  of the multiloop Baxter equation}},
  \href{https://doi.org/10.1016/j.nuclphysb.2009.10.030}{\emph{Nucl. Phys. B}
  {\bfseries 827} (2010) 565}
  [\href{https://arxiv.org/abs/0908.0520}{{\ttfamily 0908.0520}}].

\bibitem{Dokshitzer:2005bf}
Y.L.~Dokshitzer, G.~Marchesini and G.P.~Salam, \emph{{Revisiting parton
  evolution and the large-x limit}},
  \href{https://doi.org/10.1016/j.physletb.2006.02.023}{\emph{Phys. Lett. B}
  {\bfseries 634} (2006) 504}
  [\href{https://arxiv.org/abs/hep-ph/0511302}{{\ttfamily hep-ph/0511302}}].

\bibitem{Dokshitzer:2006nm}
Y.L.~Dokshitzer and G.~Marchesini, \emph{{N=4 SUSY Yang-Mills: three loops made
  simple(r)}},
  \href{https://doi.org/10.1016/j.physletb.2007.01.016}{\emph{Phys. Lett. B}
  {\bfseries 646} (2007) 189}
  [\href{https://arxiv.org/abs/hep-th/0612248}{{\ttfamily hep-th/0612248}}].

\bibitem{Velizhanin:2010cm}
V.N.~Velizhanin, \emph{{Six-Loop Anomalous Dimension of Twist-Three Operators
  in N=4 SYM}}, \href{https://doi.org/10.1007/JHEP11(2010)129}{\emph{JHEP}
  {\bfseries 11} (2010) 129} [\href{https://arxiv.org/abs/1003.4717}{{\ttfamily
  1003.4717}}].

\bibitem{Kniehl:2020rip}
B.A.~Kniehl and V.N.~Velizhanin, \emph{{Nonplanar Cusp and Transcendental
  Anomalous Dimension at Four Loops in $\mathcal{N}$=4 Supersymmetric
  Yang-Mills Theory}},
  \href{https://doi.org/10.1103/PhysRevLett.126.061603}{\emph{Phys. Rev. Lett.}
  {\bfseries 126} (2021) 061603}
  [\href{https://arxiv.org/abs/2010.13772}{{\ttfamily 2010.13772}}].

\bibitem{Kniehl:2021ysp}
B.A.~Kniehl and V.N.~Velizhanin, \emph{{Non-planar universal anomalous
  dimension of twist-two operators with general Lorentz spin at four loops in
  $N=4$ SYM theory}},
  \href{https://doi.org/10.1016/j.nuclphysb.2021.115429}{\emph{Nucl. Phys. B}
  {\bfseries 968} (2021) 115429}
  [\href{https://arxiv.org/abs/2103.16420}{{\ttfamily 2103.16420}}].

\bibitem{Moch:2017uml}
S.~Moch, B.~Ruijl, T.~Ueda, J.A.M.~Vermaseren and A.~Vogt, \emph{{Four-Loop
  Non-Singlet Splitting Functions in the Planar Limit and Beyond}},
  \href{https://doi.org/10.1007/JHEP10(2017)041}{\emph{JHEP} {\bfseries 10}
  (2017) 041} [\href{https://arxiv.org/abs/1707.08315}{{\ttfamily
  1707.08315}}].

\end{thebibliography}
\end{document}